\documentclass[iop]{emulateapj}
 \usepackage{psfrag}
 \usepackage{rotating}
 
 \shorttitle{Evolution, Mass Loss and UV Emission of $l$~Car}
 \shortauthors{Neilson et al.}
  
\begin{document}
  
 \title{The Secret Lives of Cepheids: Evolution, Mass Loss, and Ultraviolet Emission of the Long-Period Classical Cepheid \lowercase{\emph{l}} Carinae \footnote{Based on observations made with the NASA/ESA \textit{Hubble Space Telescope}, obtained at the Space Telescope Science Institute, which is operated by the Association of Universities for Research in Astronomy, Inc., under NASA contract NAS 5-26555. These observations are associated with program \#13019. This work is also based on observations obtained with \textit{XMM-Newton}, an ESA science mission with instruments and contributions directly funded by ESA Member States and the USA (NASA), associated with program \#060374.}}

 \author{Hilding R.~Neilson\altaffilmark{1}}
\author{Scott~G.~Engle\altaffilmark{2}}
\author{Edward~F.~Guinan\altaffilmark{2}}
\author{Alexandra~C.~Bisol\altaffilmark{2}}
 \author{Neil~Butterworth\altaffilmark{3}}
 \altaffiltext{1}{Department of Astronomy \& Astrophysics, University of Toronto, 50 St.~George Street, Toronto, ON, M5S~3H4, Canada}
\email{neilson@astro.utoronto.ca}
\altaffiltext{2}{Department of Astrophysics \& Planetary Science, Villanova University, 800 E.~Lancaster Ave., Villanova, PA 19085, USA}
\altaffiltext{3}{American Association of Variable Star Observers (AAVSO), 49 Bay State Rd., Cambridge, MA 02138, USA}

\begin{abstract}
The classical Cepheid $l$~Carinae is an essential calibrator of the Cepheid Leavitt Law as a rare long-period Galactic Cepheid.   Understanding the properties of this star will also constrain the physics and evolution of massive ($M \ge 8~M_\odot$) Cepheids. The challenge, however, is precisely measuring the star's pulsation period and its rate of period change. The former is important for calibrating the Leavitt Law and the latter for stellar evolution modeling. In this work, we combine previous time-series observations spanning more than a century with new observations to remeasure the pulsation period and compute the rate of period change.  We compare our new rate of period change with stellar evolution models to measure the properties of $l$~Car, but find models and observations are, at best, marginally consistent.  The results  imply that $l$~Car does not have significantly enhanced mass-loss rates like that measured for $\delta$~Cephei. We  find that the mass of $l$~Car is about 8 -- 10~$M_\odot$.  We  present Hubble Space Telescope COS observations that also differ from measurements for $\delta$~Cep, and $\beta$~Dor.  These measurements further add to the challenge of understanding the physics of Cepheids, but do hint at the possible relation between enhanced mass loss and ultraviolet emission, perhaps both due to the strength of shocks propagating in the atmospheres of Cepheids.
\end{abstract}


\section{Introduction}

The star \emph{l}~Carinae is a bright $(\langle V \rangle \approx 3.6$~mag), long-period ($P = 35.5$ day) classical Cepheid, making it one of the most important Galactic Cepheids for constraining stellar evolution models as well as calibrating the Cepheid Leavitt Law \citep{Leavitt1908}.  Long-period~Cepheids like \emph{l}~Car are uncommon in the Galaxy, so as the nearest long-period~Cepheid, at a distance of $498^{+55}_{-45}$pc \citep{Benedict2007}, long-term observations present the opportunity to probe the details of stellar pulsation and evolution.

The forthcoming era of the \emph{James Webb Space Telescope} (\textit{JWST}) will allow for measurements of the Hubble constant to a precision of about 1\% using Cepheids \citep{Freedman2012}. However, this precision requires calibrating the Leavitt Law using Large Magellanic Cloud Cepheids to measure the slope and Galactic Cepheids to anchor the zero point.  As a rare long-period Galactic Cepheid, \emph{l}~Car is uniquely important for anchoring that zero point, hence it is essential to measure the pulsation period and luminosity precisely.

Similarly, \emph{l}~Car is a powerful laboratory for constraining stellar astrophysics.  It is a massive Cepheid evolving along a short-lived stage of stellar evolution and will eventually explode as a supernova or evolve as a Super Asymptotic Giant Branch star, depending on how massive it is \citep[e.g.][]{Eldridge2004}.   Furthermore, this specific star is important for understanding the transition from red supergiant to blue supergiant and vice versa \citep{Maeder2000},  short but crucial stages of stellar evolution for understanding stellar mass loss \citep{Mackey2012, Meyer2014}, rotation \citep{Anderson2014}, supernova progenitors \citep{Georgy2012} and other physics \citep{Langer2012}.  Accurate observations and models are required to differentiate physical processes and evolutionary scenario.

As one of the nearest Cepheids, \emph{l}~Car has been observed nearly continuously for more than a century \citep{Cousins1924}.   Because of its brightness,  \emph{l}~Car has been observed from ultraviolet (UV) to infrared (IR) wavelengths.  \cite{Vitense1994} presented {\it International Ultraviolet Explorer} (IUE) satellite observations that reveals the presence of strong far-UV line emission that varies and is periodic. Recently, \cite{Engle2012} found evidence for photospheric temperatures up $\approx 10^5~$K based on HST-COS NUV and FUV that also showed periodic variations.
 At IR wavelengths, \cite{Kervella2009} detected a circumstellar envelope at a distance $\approx 10 - 100$~AU from the star, possibly arising from significant mass loss. This Cepheid has also been observed using interferometry to measure its angular diameter, $\theta = 2.992\pm 0.012$~mas \citep{Kervella2004a}.  Combined with HST FGS parallax measurements \citep{Benedict2007}, the  distance is $498^{+55}_{-45}$~pc and hence mean radius is $R = 159.9 \pm 16.6~R_\odot$.  These observations are necessary for constraining the fundamental properties of \emph{l}~Car while also presenting new mysteries about its atmosphere, circumsteller medium and about Cepheids in general.

To understand these observations better, we require precise measurements of the pulsation period.  Because \emph{l}~Car is a massive star that evolves on short time scales, the pulsation period should also be observed to change. \cite{Cousins1924} presented period measurements of the star from 1891 to 1924 and found that the period changed from $P = 35.5236~$days increasing to $P = 35.57~$days by 1924.  Further period studies have been since conducted, but results have been ambiguous \citep[e.g.][]{Feinstein1969, Cogan1980, Kervella2004a}.
However, continuous and precise period measurements for \emph{l}~Car are an ongoing challenge because of the star's long period relative to other nearby classical Cepheids, hence requiring months of observations to determine reliable timing measurements.  Also, \emph{l}~Car is too bright to be observed using conventional CCD photometry.  These two challenges hinder a thorough period study of this Cepheid.

Precise period determinations and measurements of period change are critical for constraining the evolution of Cepheid variable stars \citep{Eddington1918}. \cite{Turner2006} compiled rates of period change for about 200 Galactic Cepheids, including \emph{l}~Car, and showed that period change constrains Cepheid evolution models and indicates which crossing of the instability strip that a Cepheid is evolving.  Similarly, \cite{Neilson2012a} and \cite{Neilson2014b} compared the measured rate of period change for the short-period Cepheid Polaris  with stellar evolution models to constrain its wind and mass-loss properties. \cite{Neilson2012b} computed population synthesis models of Cepheids with the \cite{Turner2006} sample and found that enhanced mass-loss (of the order $10^{-7}~M_\odot$~yr$^{-1}$) is a common phenomena.  However, because of the rarity of massive Cepheids relative to less massive Cepheids, population synthesis models do not significantly constrain the mass-loss rates of massive Cepheids like \emph{l}~Car and RS~Pup.  More recently, \cite{Anderson2014} compared evolution models with stellar rotation to show that Cepheids may also be significantly rotating, adding yet another ingredient.

While these period changes test stellar evolution models, recent \emph{Kepler} observations of V1154~Cyg suggest random-like period variations of the order of  $\Delta P \approx 0.01 P$.  This phenomena has also been detected in two other Cepheids \citep{Evans2015}. This period jitter could be due to period instabilities \citep[e.g.][]{Poleski2008} or by convective cells on the surface of the star \citep{Neilson2014}.  \cite{Anderson2015} presented new interferometric and spectroscopic measurements that show radial velocity jitter where the radial velocity curve varies cycle-to-cycle.  This is analogous to the photometric jitter measurements. If the latter hypothesis is true, then the pulsation period of \emph{l}~Car could intrinsically vary by at least 0.03~day or more, further complicating the picture.

In this work, we compile and reanalyze published data for \emph{l}~Car, which we complement with new observations to measure its pulsation period and rate of period change.  We present our data in Sect.~2 spanning from 1871 to 2012 and describe the period determination and measured period change. In Sect.~3, we compute new stellar evolution models to compare to our measured rate of period change. New ultraviolet spectral observations are presented and discussed in Sect.~4.  These results are summarized and discussed in Sect.~5.

\section{Photometric Observations of \lowercase{\emph{l}}~Car}

To measure the period and rate of period change for \emph{l}~Car, we use published data spanning from 1871 to 1990 which we complement with new observations for the year 2012 taken by one of us (Butterworth).  We measure new times of maximum light for period determinations from the data when the light curve data is available, otherwise we use the published timings of maximum light.  A sample of the data we use is listed in Table~\ref{t1}, the entire table is available in machine-readable format.

\begin{table*}[t]
\caption{A sample of published and new timing data for \emph{l}~Car}
\begin{center}
\begin{tabular}{ccccl}
\hline
\hline
Timing Maximum (HJD) &Epoch & O-C(days) & Weighting& Source  \\
\hline
2404640.625&	-932&	7.8177&	0.5&	Innes (1897)\tablenotemark{*}\\
2404671.625&	-931&	3.1982&	0.5&	Innes (1897)\tablenotemark{*}\\
2404921.625&	-924&	4.6196&	0.5&	Innes (1897)\tablenotemark{*}\\
2411846.417&	-729&	0.0000&	0.5&	Innes (1897)\tablenotemark{*}\\
2412237.417&	-718&	0.0000&	0.5&	Innes (1897)\tablenotemark{*}\\
2413235.417&	-690&	3.1982&	0.5&	Innes (1897)\tablenotemark{*}\\
2413267.417&	-689&	-0.3554&	0.5&	Innes (1897)\tablenotemark{*}\\
2413305.417&	-688&	2.1321&	0.5&	Innes (1897)\tablenotemark{*}\\
2413553.417&	-681&	1.4214&	0.5&	Innes (1897)\tablenotemark{*}\\
2413587.417&	-680&	0.0000&	0.5&	Innes (1897)\tablenotemark{*}\\
2413658.417&	-678&	0.0000&	0.5&	Innes (1897)\tablenotemark{*}\\
2413694.417&	-677&	0.0000&	0.5&	Innes (1897)\tablenotemark{*}\\
2413730.417&	-676&	0.7107&	0.5&	Innes (1897)\tablenotemark{*}\\
2413766.417&	-675&	1.0661&	0.5&	Innes (1897)\tablenotemark{*}\\
2413870.417&	-672&	-1.4214&	0.5&	Innes (1897)\tablenotemark{*}\\
2413906.417&	-671&	-1.0661&	0.5&	Innes (1897)\tablenotemark{*}\\
2422186.980&	-438&	-0.1900&	2.0&	\cite{Berdnikov2003}\\
2422187.200&	-438&	0.0300&	2.0&	Berdnikov et al. (2003)\\
2422968.320&	-416&	-0.6200&	2.0&	Berdnikov et al. (2003)\\
2423964.130&	-388&	0.2100	&2.0&	Berdnikov et al. (2003)\\
\hline
\end{tabular}
\label{t1}
\tablenotetext{*}{Please see Section 2 for a discussion of the times of maximum reported by \cite{Innes1897}.}
\end{center}
\end{table*}

\begin{figure*}[t]
\begin{center}
\plotone{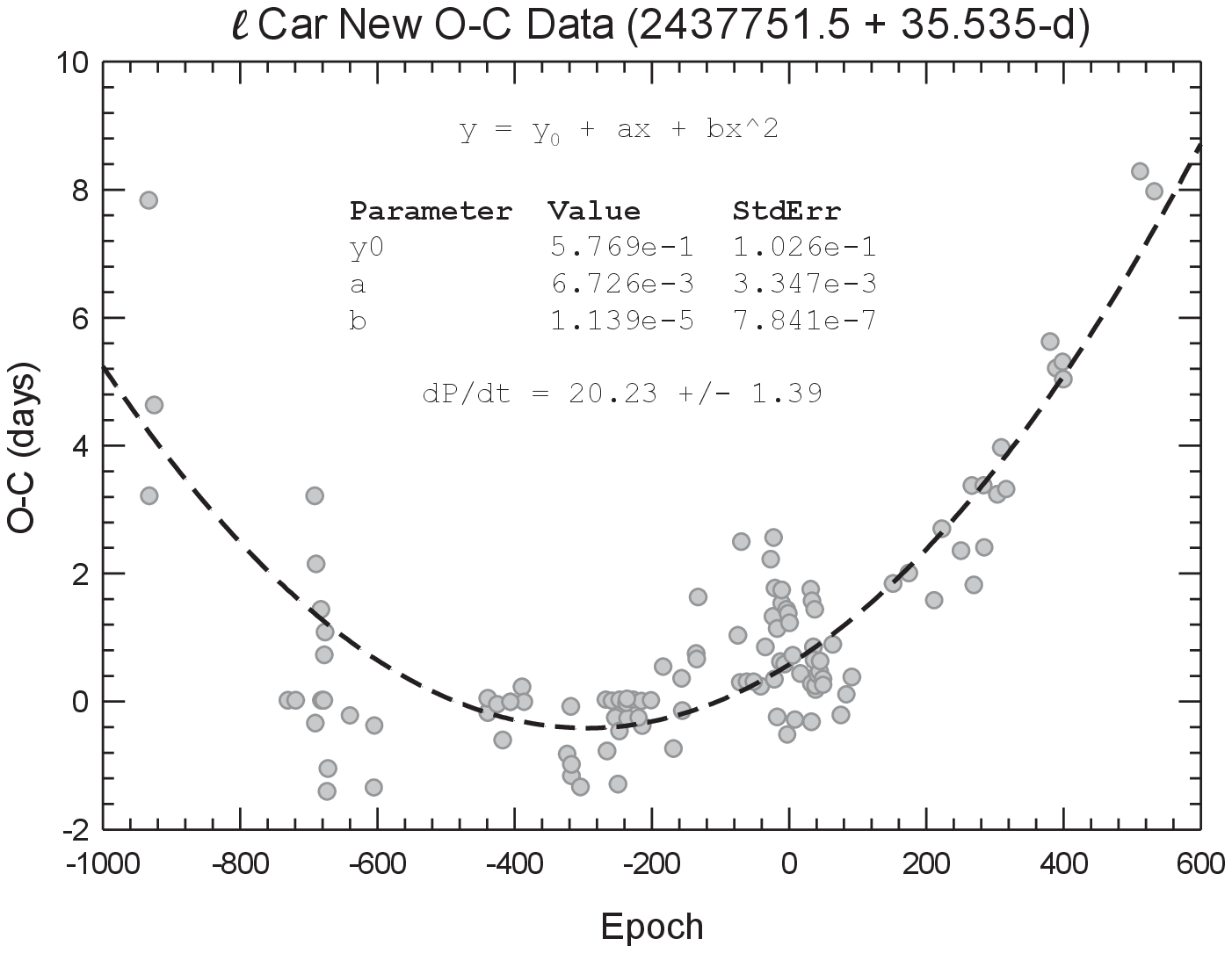}
\end{center}
\caption{O-C diagram for a century of observations of \emph{l}~Car, along with the best-fit parabolic function (dashed line). }\label{f1}
\end{figure*}

We fit available light curve data using a Fourier series fit to calculate new timing maxima.  That timing maxima  is combined with the \cite{Roberts1901} and \cite{Cousins1924} timing measurements to produce the observed-calculated (O-C) diagram.  We assume the ephemeris period from \cite{Feinstein1969} to be consistent with other works.  We plot the O-C diagram in Fig.~\ref{f1} and fit the data with a quadratic function to measure the secular rate of period change \citep[e.g.][]{Percy2008}.  The fit is computed with more weight given if the light curve data is available for this work and less statistical weight for the older \cite{Innes1897}, \cite{Roberts1901} and \cite{Cousins1924} data (see Table 1). We note that \cite{Innes1897} only published the dates of observed times of maximum, but not the actual times. For the first three times of maximum, the JD published by Innes is adjusted to represent local midnight in Cordoba, Argentina, where those observations were made. For the rest of the Innes timings, adjustments were made to local midnight for Cape Town, South Africa. These timings, although important as they are the earliest found for \emph{l}~Car, are also given a low weight (see Table 1), and the error on each observation is on the order of $\pm$2 days. 


Based on the timing data and the O-C diagram, we measure the rate of period change for \emph{l}~Car to be $\dot{P} = +20.23\pm  1.39$~s~yr$^{-1}$. This rate is somewhat different from that reported by \cite{Turner2010}, $\dot{P} = +26.36$~s~yr$^{-1}$.  We believe the smaller period change value found in this study is due to the inclusion of very early times of maximum, in addition to the most recent data not available for the earlier studies. \cite{Breitfelder2016} presented another measurement of the rate of period change for \emph{l}~Car using the SPIPS method \citep{Merand2015}.  Their measured rate, $\dot{P} = 27.283\pm 0.984~$s~yr$^{-1}$, is consistent with  that measured by \cite{Turner2010} and employs a shorter time span of timing measurements than our analysis. If we choose to leave out the timings earlier than Epoch = $-$600, the resulting fit gives $\dot{P} = +23.21\pm  2.19$~s~yr$^{-1}$, bringing the value closer to that reported in \cite{Turner2010} and \cite{Breitfelder2016}. However, despite the larger uncertainty of some of these points (discussed earlier), we still believe they should be included in analysis.   We also test the data for changes in the brightness amplitude, but over the past century there is no evidence that the amplitude has changed by more than one-tenth of a magnitude.  There are hints of smaller scale amplitude changes but, due to the spans of time between complete, calibrated lightcurves, no definitive amplitude changes are found.  
These results suggest that \emph{l}~Car is evolving on the third crossing of the instability strip since its period change is both positive and relatively small for a star of its mass. We compare this rate of period change with theoretical rates from stellar evolution models in the next section.

\section{Theory vs Period Change Observations}
We compute stellar evolution models to compare with the observed fundamental parameters for \emph{l}~Car using the \cite{Yoon2005} code \citep[see][for details]{Neilson2011, Neilson2012a, Neilson2012b, Neilson2014b}.  We assume the same parameters for overshooting and composition as in \cite{Neilson2014b}. The overshooting parameter is chosen to be consistent with evolutionary model fits to the eclipsing binary Cepheid OGLE-LMC-CEP0226 \citep{Cassisi2011}  and compute models for masses ranging from  $M = 3$ to $M = 11.5~M_\odot$.  Stellar evolution models with masses $M > 11.5~M_\odot$ do not form Cepheid blue loops because of the assumed overshooting parameter.  The structure of the Cepheid blue loop is an ongoing challenge for stellar evolution theory \citep{Eldridge2015}. 

 Period change predictions are computed analytically using period-mean density relation $P\sqrt{\rho} = Q$ and taking the time derivative.  This is the same method as employed by \cite{Turner2006, Neilson2012a, Neilson2012b} and \cite{Neilson2014b}. Using this method, we compute the {\it relative} change of period $\dot{P}/P$ so that we can avoid any errors in directly modelling the pulsation period using linear and non-linear pulsation models.

We compare our models to measured fundamental stellar parameters.  We compute the radius is $R = 159.9 \pm 16.6~R_\odot$ from angular diameter and parallax measurements. Further, \cite{Kervella2009} measured an effective temperature, $T_{\rm{eff}} = 4860 \pm 150~$K, yielding a stellar luminosity, $\log (L/L_\odot) = 4.107_{-0.174}^{+0.124}$.  
\begin{figure*}[t]
\begin{center}
\plottwo{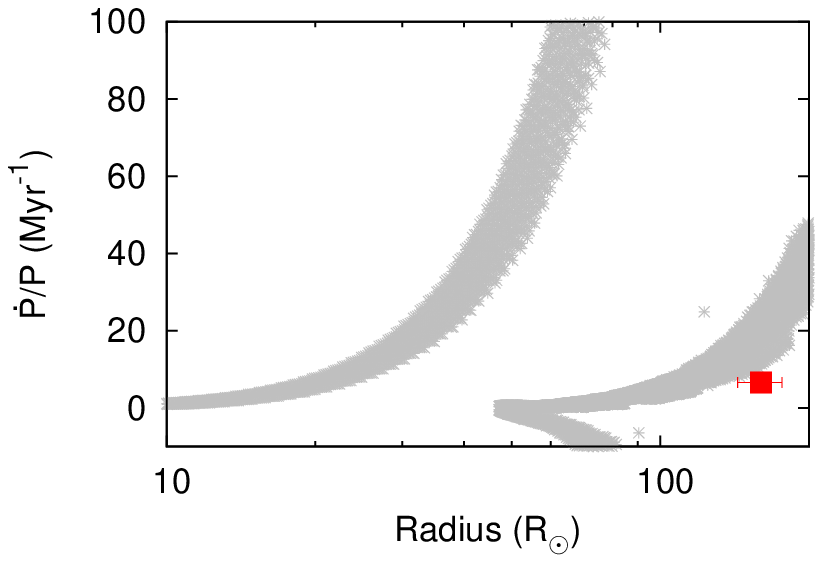}{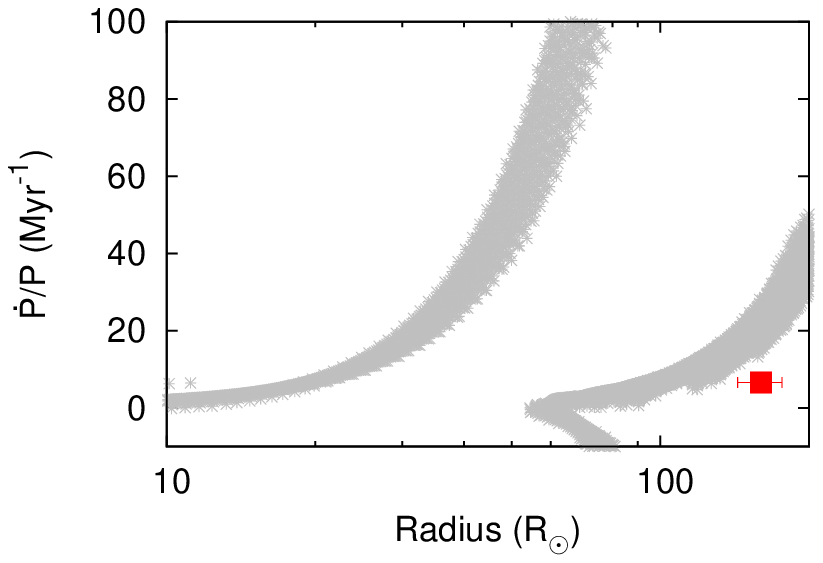}
\end{center}
\caption{Theoretical rates of period change computed from our stellar evolution models assuming Cepheid mass-loss rates $10^{-9}$ (left) and $10^{-6}~M_\odot~$yr$^{-1}$ (right). The red square represents the measured period change and radius for \emph{l}~Car.}\label{f2}
\end{figure*}

We present rates of period change as a function of radius in Fig.~\ref{f2} for two cases: models assuming a Cepheid mass-loss rate of $10^{-9}~M_\odot~$yr$^{-1}$ or   $10^{-6}~M_\odot~$yr$^{-1}$.  Each plot shows three different regimes for period change. Models with large positive rates of period change represent Cepheid evolving along the first crossing of the instability strip, models with negative rates of period the second crossing and those with small positive period change the third crossing. Based on the plots, we conclude that the rate of period change for \emph{l}~Car is inconsistent with large mass-loss rates such that $\dot{M} < 10^{-7}~M_\odot$~yr$^{-1}$.  This mass-loss rate is consistent with the results of \cite{Kervella2009}, but is smaller than mass-loss rates measured for the short-period Cepheid Polaris \citep{Neilson2012a, Neilson2014b} and the prototype $\delta$~Cephei \citep{Marengo2010, Matthews2012}.  We confirm that \emph{l}~Car is evolving along the third crossing of the instability strip.

We take this analysis a step further and correlate our predicted fundamental stellar parameters with the measured radius and rate of period change.  For models with assumed mass-loss rates of $10^{-6}$ and $10^{-7}~M_\odot$~yr$^{-1}$, no models agree with the measured radius and period change. However, for mass-loss rates $< 10^{-7}~M_\odot$~yr$^{-1}$, we measure fundamental stellar parameters from our stellar evolution models for period changes within $4\sigma$ of our measured value.  These parameters are  presented in Tab.~\ref{t2}. The predicted ages and luminosities are correlated with stellar mass, hence more massive models are both brighter and older than less massive models.

Another possibility is that the mean radius of $l$~Car is overestimated.  Because interferometric observations measure a limb-darkening angular diameter and not the actual angular diameter as defined by models then the radius could be overestimated by a few percent \citep{Merand2015}. If the radius is indeed overestimated, then the overlap between models and observations is much obvious and suggests that the mass of $l$~Car can be up to 10~$M_\odot$, but the predicted effective temperatures ceases to agree with observations.  It is unclear whether the differences between models and observation due to challenges with observations or errors in our models.

\begin{table}[t]
\caption{Fundamental parameters consistent with measured period change and radius}
\label{t2}
\begin{center}
\begin{tabular}{lc}
\hline
\hline
Fundamental Parameter & Value \\
\hline
Age (Myr)  & $31.25\pm 1.05$ \\
Mass ($M_\odot$) & $8.66\pm 0.14$ \\
$T_{\rm{eff}}$~(K) & $5000\pm 60$ \\
$\log~L/L_\odot$ & $4.06 \pm 0.03$ \\
\hline
\end{tabular}
\end{center}
\end{table}

\begin{figure*}[t]
\begin{center}
\plotone{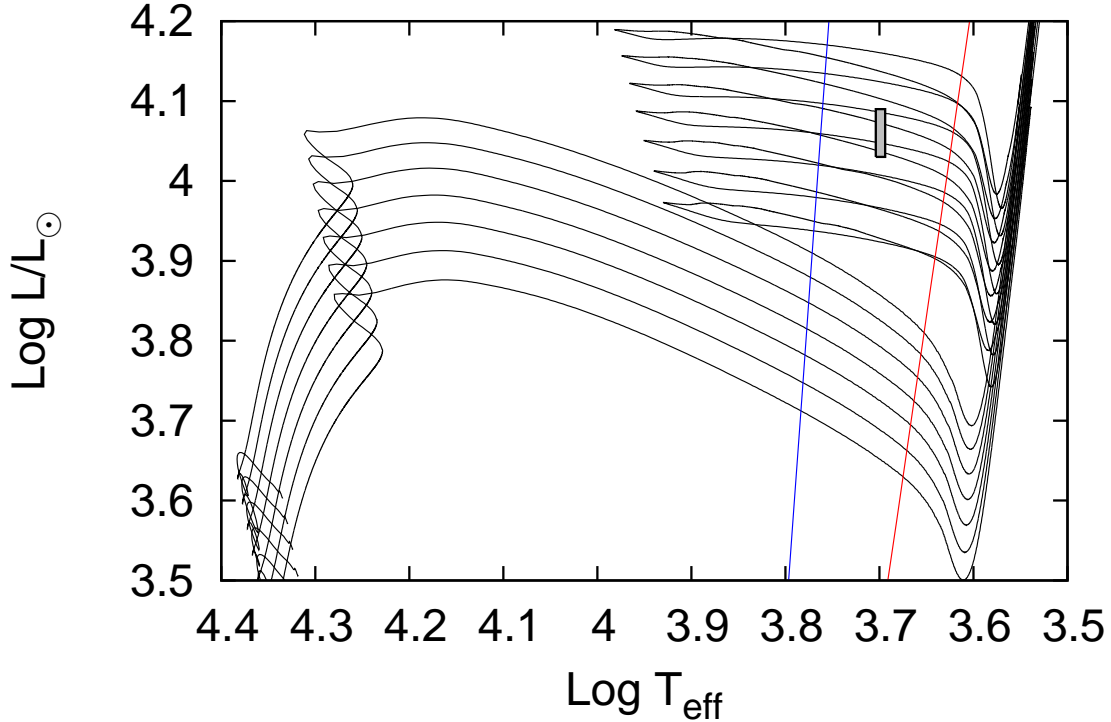}
\end{center}
\caption{Hertzsprung-Russell diagram for stellar evolution with initial masses, $M = 8$ -- $9.2~M_\odot$ in steps of $0.2~M_\odot$ (black lines) plotted with our measured effective temperature and luminosity (grey box).  The blue and red lines represent the blue and red edges of the Cepheid instability strip \citep{Bono2000}, respectively.}\label{f3}
\end{figure*}

We plot the Hertzsprung-Russell diagram for stellar evolution models with initial masses ranging from $M = 8.5~M_\odot$ to $8.8~M_\odot$ that intersect our best-fit stellar properties in Fig.~\ref{f3}. This mass is much smaller than previously considered; \cite{Caputo2005} determined a stellar evolutionary mass $M \approx 12.5~M_\odot$.  \cite{Kervella2009} noted that the mass of \emph{l}~Car is about 13~$M_\odot$.  However, if we consider the period-mass-radius relation  \citep{Fricke1972, Gieren1989} then $M \approx 7.5~M_\odot$, which is closer to what we predict here. We also note that the measured mass and luminosity depend on the assumed convective core overshooting parameter; for instance ignoring overshooting during main sequence evolution leads to a predicted mass that is about 10 -- 20\% larger. While the predicted mass differs from previous results, our predicted effective temperature and luminosity are consistent.

We also determine the stellar age to be about $31~$Myr, but the real question about \emph{l}~Car is how long will it remain a classical Cepheid.  Using the predicted red edge of the instability strip as a limit we find that \emph{l}~Car will cease being a Cepheid in less than 31,000 years;  relative to the third crossing time scale of about 0.1~Myr for an 8--9~$M_\odot$ Cepheid. 

\section{Observation at Ultraviolet and X-ray Wavelengths}
\begin{figure*}[t]
\begin{center}\label{figX}
\epsscale{1}
\plotone{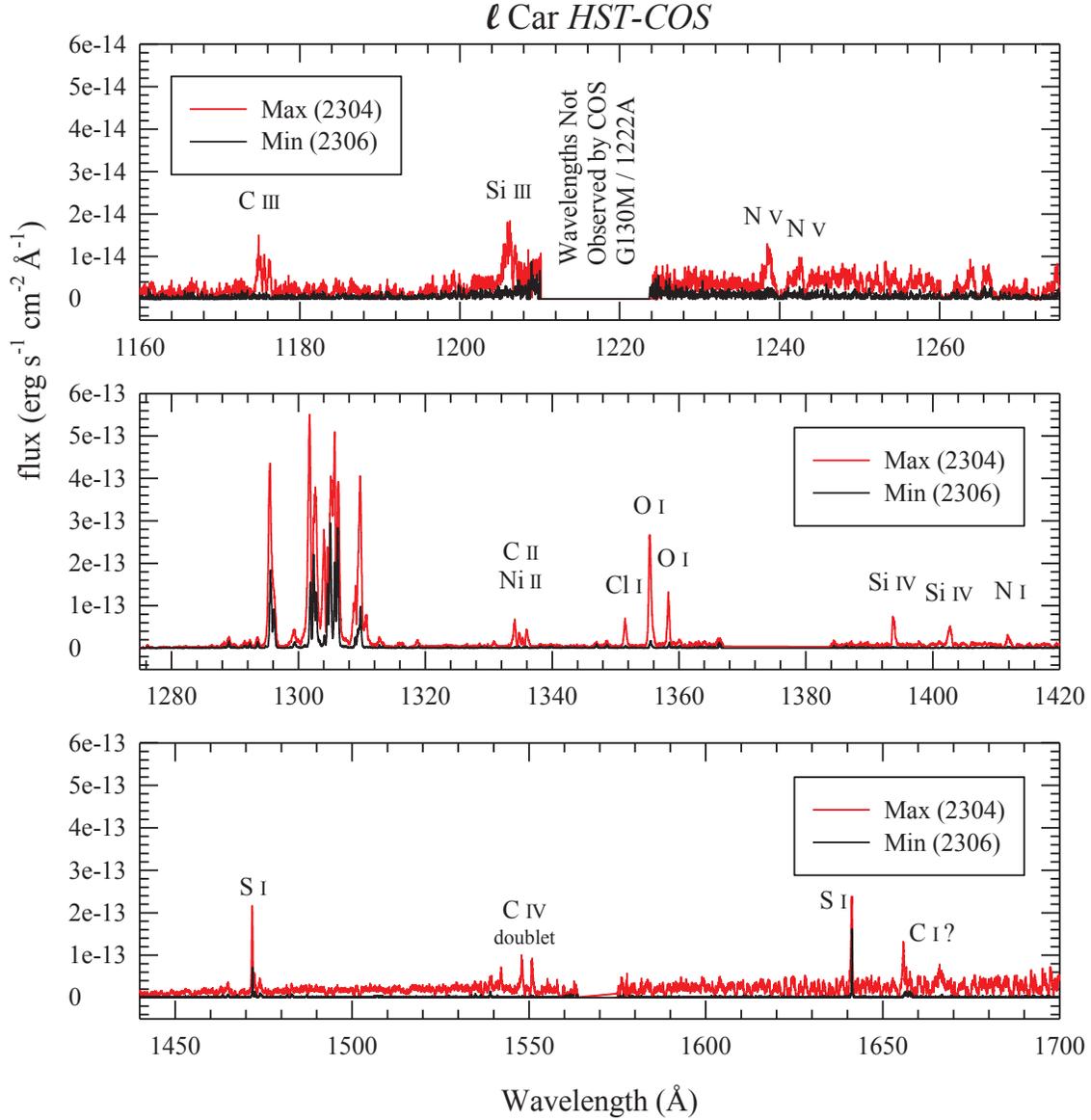}
\end{center}
\caption{HST-COS ultraviolet spectra  for phases with maximum (red lines) and minimum (black lines) Emission is significantly stronger at maximum suggestive of significant photospheric heating. Note that the middle and lower panels are set to the same y-axis scale, but the scale of the top panel has been magnified 10$\times$ due to the relative weakness of these shorter-wavelength emission lines.}
\end{figure*}

Six observations of \emph{l}~Car were carried out in 2012 -- 2013 with the Cosmic Origins Spectrograph onboard the Hubble Space Telescope (HST-COS). The goal of these observations was to further the work of previous International Ultraviolet Explorer (IUE) studies of \emph{l}~Car. \cite{Vitense1994} studied several IUE spectra of \emph{l}~Car, investigating emission line fluxes and profiles at various phases. They found the Cepheid's UV emission lines to undergo phased variability, and further concluded there was ``probably [...] evidence of mass loss.'' The HST-COS wavelength region ($\approx1150 - 1750~$\AA) is excellent for such studies as it covers numerous emission line features from atmospheric plasmas of $\approx10,000-300,000$~K (see Fig.~\ref{figX}).

\begin{figure*}[t]
\begin{center}
\epsscale{0.8}
\plotone{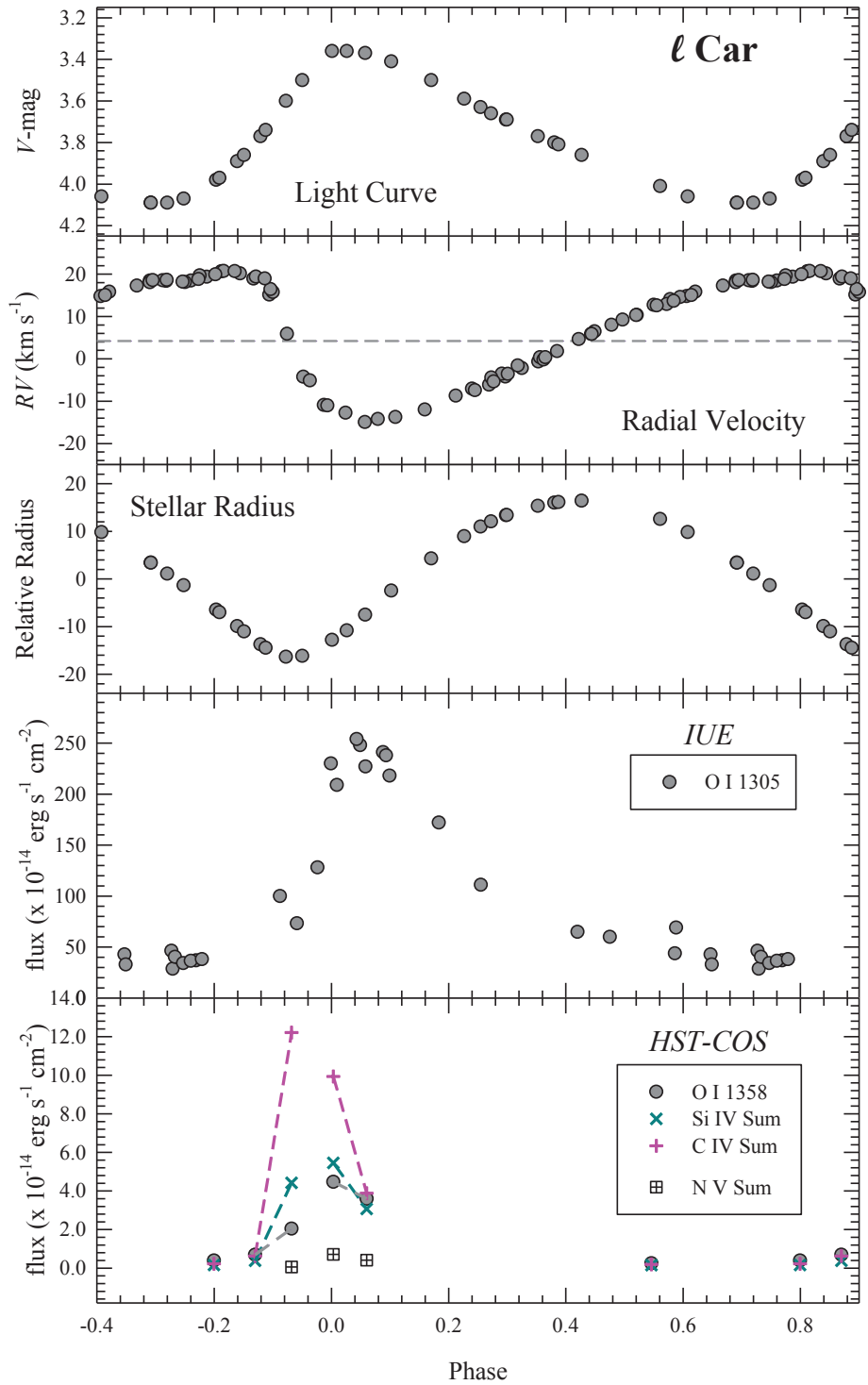}
\end{center}
\caption{Variation of the different measured properties of \emph{l}~Car as a function of pulsation phase are shown. The top panels display the variations of V-band brightness, radial velocity and stellar radius plotted against phase. The bottom two panels show the FUV emission line variability from IUE and HST-COS observations. The integrated fluxes for the O~\textsc{i}~1305 triplet emission were taken by \cite{Vitense1994}, carried out with the IUE satellite during the early 1990s. The strong O~\textsc{i}~1305 emission blend is composed of resonance emissions (in part) pumped by H~\textsc{i} Ly-$\beta$ 1028\AA~emission \citep{Koncewicz2007}. The O~\textsc{i}~1305 triplet emission reaches a maximum near 0.05$\phi$. As shown in the lower panel, it is notable that the FUV emission fluxes reach maximum strengths between 0.95 -- 0.05$\phi$. Although based on fewer observations from HST-COS, the C~\textsc{iv}~1550 emission peaks near 0.95$|phi$, while O~\textsc{i}~1358, Si~\textsc{iv}~1400 and N~\textsc{v}~1240 emission fluxes reach maximum strengths near the star's maximum optical brightness at 0.0$\phi$. These FUV emissions appear to arise from a pulsation-induced shock traveling outward from the star.}\label{figY}
\end{figure*}

As shown in Fig.~\ref{figY}, the HST-COS observations confirm the previous \cite{Vitense1994} IUE findings of variable, pulsation-phased UV emission line activity. The HST-COS observations generally show similar behavior to the UV emission line variations of other Cepheids \citep[see][]{Engle2014, Engle2015}. The strong O \textsc{i} 1305 triplet (blended with smaller contributions from S~\textsc{i} and Si~\textsc{ii}) emission flux from \cite{Vitense1994} are plotted in the figure. The O~\textsc{i}~1305 emission reaches maximum strength near $\sim$0.05$\phi$. However, the O~\textsc{i}~1305 triplet emission is a resonance spectral feature (in part) pumped by H~\textsc{i}~Ly-$\beta$ at 1028\AA~\citep{Koncewicz2007}. As shown in the figure, the integrated UV emission line fluxes begin increasing after 0.9$\phi$, as the Cepheid is approaching minimum stellar radius (while the photosphere is shrinking, but decelerating). This is the phase range where a pulsation-induced shock is expected to emerge from the photosphere and propagate through the outer atmosphere. The abruptness of the line flux increases agrees with a sudden heating mechanism, such as a shock. 

In addition to the O~\textsc{i}~1305 emission feature, the emission line fluxes from several other representative FUV lines secured with HST-COS are given in Fig.~\ref{figY}.  Most of these lines originate from hotter plasma than the O~\textsc{i}~1305 emission. The emission line fluxes of the C~\textsc{iv}~1548/1550 doublet, N~\textsc{v}~1238/42 doublet, O~\textsc{i}~1358 line and Si~\textsc{iv}~1393/1403 doublet are plotted in the figure. These lines could not be accurately measured previously with IUE. Except for the O~\textsc{i}~1358 line, the lines originate from hot plasmas with T $> 50 \times 10^4$ K. It appears that the C~\textsc{iv} emission peaks first near $\sim$ 0.93--0.95$\phi$. But there are not sufficient observations to precisely define the phase. The Si~\textsc{iv}~1400 and O~\textsc{i}~1358 line fluxes reach maximum strengths somewhat later at $\sim$0.0$\phi$. As shown, the N~\textsc{v} emission is very weak and is only definitely detected near maximum light. This emission line is too weak to be unambiguously measured at other phases even with HST. From \cite{Engle2014}, the approximate plasma temperatures of the various emission lines are: O~\textsc{i}~1305 ($\sim 1 - 2 \times 10^4$ K), O~\textsc{i}~1358 ($\sim 1.5 - 2.0 \times 10^4$ K), Si~\textsc{iv}~1400 ($\sim 50 - 70 \times 10^4$ K), C~\textsc{iv}~1550 ($\sim 80 - 150 \times 10^4$ K) and N~\textsc{v}~1240 ($\sim 250 - 300 \times 10^4$ K).

However, there is interestingly one aspect where \emph{l}~Car behaves differently from $\delta$~Cep and $\beta$~Dor \citep[see][]{Engle2014, Engle2015}. As said, the Cepheid shows pulsation phase-dependent variations in UV emission line strengths (see Fig.~\ref{figY}). However, the N~\textsc{v}~1239/1243~\AA~doublet (peak formation temperature of $\sim1.5-2.5\times10^{5}$ K), which is present in all spectra of $\delta$~Cep and $\beta$~Dor, is only detected in the two of the most active spectra of \emph{l}~Car. This is interesting since the N~\textsc{v} wavelength region is essentially free of photospheric continuum flux from the cool supergiant. Therefore, even weak N~\textsc{v} emissions should be detectable. This implies that \emph{l}~Car possesses a relatively cooler outer atmosphere (at least during the majority of each pulsation cycle) compared to either $\delta$~Cep or $\beta$~Dor, and it is only during the phases of shock-enhanced emissions that the atmospheric plasmas are sufficiently heated to result in N~\textsc{v} emissions. Because \emph{l}~Car has a longer period than either $\delta$~Cep (5.6~d) or $\beta$~Dor (9.8~d), the interval between pulsation-induced shock-heating events is much longer and the shock-heated plasma has time to cool.

\begin{figure*}[t]
\begin{center}\label{figZ}
\epsscale{1}
\plotone{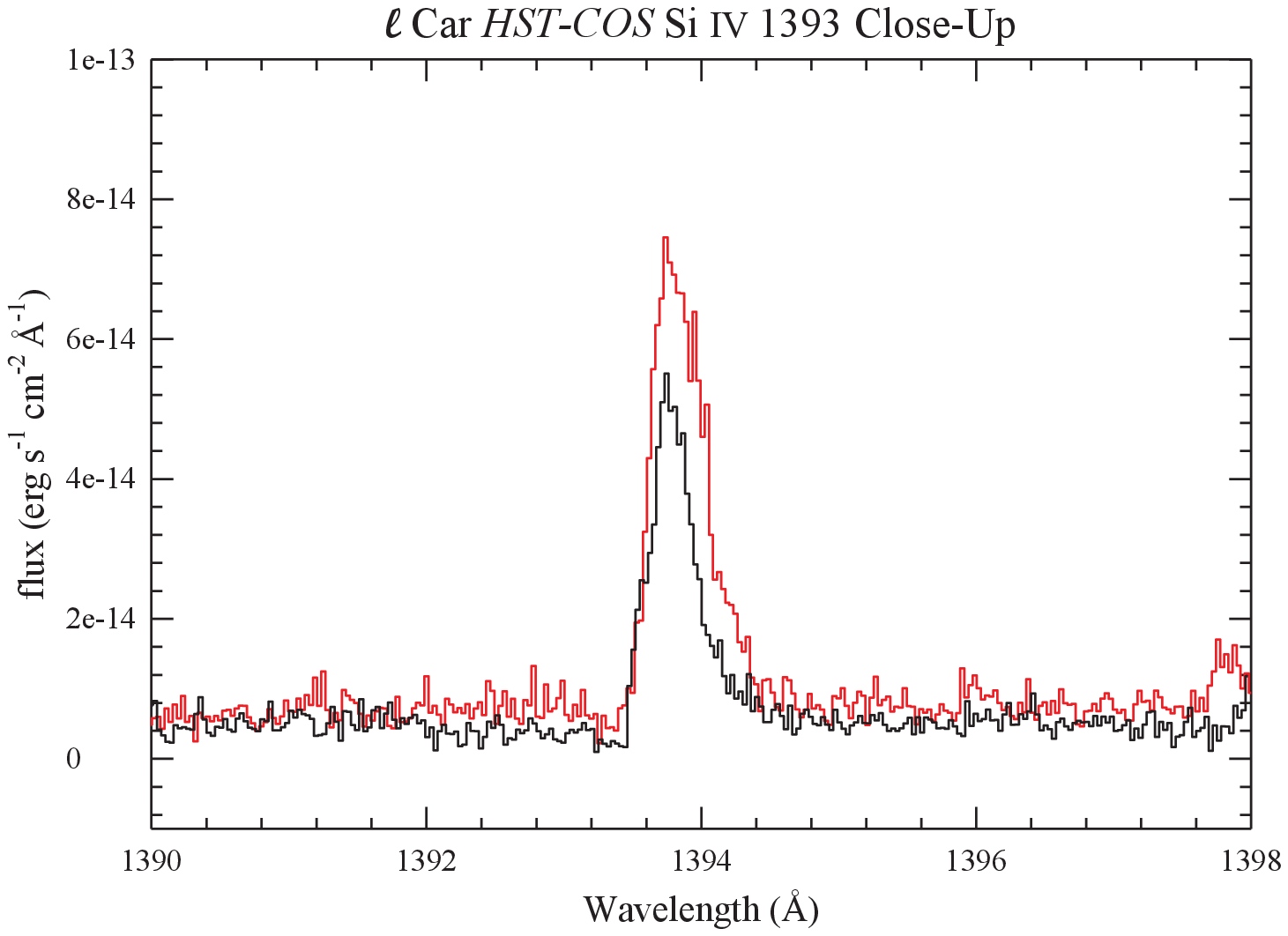}
\end{center}
\caption{A close-up of the Si~\textsc{IV}~1393 emission line of \emph{l}~Car from Fig.~\ref{figX} showing that the line is both stronger and broadened to longer wavelengths at peak emission, hinting at potential mass-loss physics.}
\end{figure*}

Additionally, there is potential evidence of mass loss in the form of blue-shifted absorption features (see Fig.~\ref{figZ}), indicating a stellar wind. At present, these ``wind features'' only show up in the more active spectra. This would imply pulsation-related variable mass loss. However, the lower signal-to-noise of the emission features in less-active (weaker emission line) spectra make it difficult to definitively identify wind signatures (such as P~Cygni features and emission line asymmetries) at those phases. Until a reliable circumstellar model is constructed and applied to the spectra, we presently report the possible spectroscopic detection of mass loss, but likely below $10^{-8}~M_\odot~$yr$^{-1}$, consistent with the results using period change measurements.

Also as part of \emph{The Secret Lives of Cepheids program}, an X-ray observation of \emph{l}~Car was carried out with XMM-Newton in February, 2010. This observation was obtained to search for evidence of X-ray emission that has been found in a small number of Cepheids, including $\delta$~Cep and $\beta$~Dor \citep{Engle2014, Engle2015}. However, \emph{l}~Car was not detected. An upper limit for its X-ray Luminosity ($L_{\rm{X}}$) was determined, using the background count rate at the position of \emph{l}~Car in the XMM data.  To produce twice this rate, given the exposure time at the position of \emph{l}~Car ($\approx47,500~$s) and the distance, $498^{+55}_{-45}$~pc \citep{Benedict2007}, \emph{l}~Car would require an X-ray luminosity of $\log (L_{\rm{X}} \approx 29.6$. The only Cepheid to have been detected at such an activity level is V473 Lyr \citep{Evans2016}. However, there is also the pulsation phase of the detection to consider. The two Cepheids having sufficient X-ray phase coverage ($\delta$~Cep and $\beta$~Dor), have phased-variable X-ray emissions that peak near a phase of 0.5 \citep{Engle2014, Engle2015}. Our XMM observation of \emph{l}~Car was secured near phase $= 0.2$, where the X-ray activity of the Cepheid could easily have been below the detection threshold of our observation.

\section{Discussion}
In this work, we analyzed almost a century of observations of the massive Cepheid \emph{l}~Car to measure a rate of period change $\dot{P} =  +20.23\pm  1.39~$s~yr$^{-1}$, a value much smaller than previously presented \citep{Turner2006}.  Much of this difference is due to our new timing measurements and reanalysis of previously observed light curves.  We also tested the data for changes in the brightness amplitude that would indicate that \emph{l}~Car is transitioning from a classical Cepheid to a red giant / supergiant star, however, no change in amplitude was detected. New UV observations were presented, offering new insights into the Cepheid wind that is not as strong as that for $\delta$~Cep or $\beta$~Dor.

We compared our new rate of period change plus measured mean radius to stellar evolution models, verifying the hypothesis that \emph{l}~Car is evolving along the third crossing of the Cepheid instability strip. Based on these models, we predict fundamental stellar parameters for the star, including its luminosity and mass, along with its mass-loss rate and age.  The best-fit mass is smaller than derived from previous stellar evolution model fits to effective temperature and luminosity and smaller than masses measured from pulsation relations \citep{Caputo2005}. Furthermore, we compute the age of \emph{l}~Car to be about 31~Myr, and that \emph{l}~Car will cease being a classical Cepheid in less than 31,000 years, even though the uncertainty in age is about one million years.  This is because the predicted age is correlated with the predicted mass, but not the time scale for evolution to the red edge of the instability strip.

It should be noted that our results suggest that \emph{l}~Car is evolving closer to the middle of the instability strip than to the red edge \citep{Luck2008, Kervella2009}.   But, if we compare the fundamental parameters of those models with an adiabatic period-mass-radius relation \cite{Gieren1989}, then the period is consistent with that of \emph{l}~Car for the coolest evolution models, again verifying that \emph{l}~Car is much closer to the red edge of the instability strip than to the blue edge.

While the evolutionary models provide reasonable estimates for the stellar luminosity and mass, they appear to do so only for mass-loss rates $< 10^{-7}~M_\odot$~yr$^{-1}$.  These mass-loss rates appear small, especially when compared to mass-loss rates for the less massive Cepheids $\delta$ Cep and Polaris, which have inferred mass-loss rates, $\dot{M} \ge 10^{-7}~M_\odot~$yr$^{-1}$.  However, the mass-loss rate is consistent with that measured from infrared observations \citep{Kervella2009}. Either way this result has implications for understanding the Cepheid mass discrepancy and Cepheid mass loss in general. \cite{Neilson2012b} suggested that enhanced mass loss is necessary to explain the observed number of Cepheids with positive period change relative to the number with negative period change.  This smaller mass-loss rate would appear to contradict that argument.   However, the \cite{Neilson2012b} result was based on population synthesis models of Cepheids that are more sensitive to smaller mass Cepheids since these stars have much longer Cepheid life times, therefore that analysis does not necessarily require \emph{l}~Car to have enhanced mass loss.  

It should be noted that it is increasingly difficult to compare measured rates of period change with stellar evolution models.  Measurement of period change require high-precision timings for observations spanning upwards of a century. Another challenge is interpretation.  For instance, both \cite{Neilson2012a} and \cite{Yel2015} found similar results comparing the measured rate of period change for Polaris with models and reported the opposite conclusions.  \cite{Neilson2014b} found that Polaris is evolving on the third crossing of the instability strip, pulsates in the first-overtone mode and is at a distance $> 115~$pc, whereas \cite{Yel2015} argued Polaris is evolving on the first crossing, is a fundamental mode pulsator and has a distance $d\approx 107$~pc.  The source of this difference is that the measured rate of period change is not consistent with standard models at any crossing, similar to the result found here.  This suggests there may be missing physics in the stellar evolution models that directly impacts the predicted rates of period change.

While the ultraviolet spectra does not show features similar to that observed in $\delta$~Cep \citep{Engle2014} but does show weaker emission features in the near-UV suggestive of plasma with temperatures about $30,000~$K.  If these features are created by shocks propagating in the stellar envelop, it is unlikely these shocks would be related to an enhanced wind as suggested for the shorter-period Cepheids \citep{Neilson2008}. One possible explanation is that $l$~Car has a ``failed'' wind.  Pulsation in the star levitates material and as that material falls back during the pulsation cycle, shocks in the following cycle interacts with the material before the material had fallen back to its original radius.  Over many pulsation cycles, that layer reaches the escape velocity and ejected as a slow wind. However, when the pulsation period is long relative to the free-fall time of that layer, that layer will not be ejected, but rains back down onto the stellar envelop as a shock-generating collision. 

The smaller mass-loss rate is consistent with the pulsation-driven wind scenario where pulsation-generated shocks enhance a stellar wind by many orders of magnitude \citep{Neilson2008, Neilson2009}. In that situation, pulsation-driven shocks are far more efficient at driving mass loss in Cepheids nearer the blue edge of the instability strip and for Cepheids with less extended photospheres, i.e., smaller-mass, shorter-period Cepheids.  This pulsation-driven mass-loss mechanism would be inefficient in \emph{l}~Car or even damped by the convection that is becoming more efficient as \emph{l}~Car evolves red ward. As the convective layers grow, pulsation-generated shocks will interact and become dampened in the convective layers, inhibiting a strong variable wind.  This would also damp any P~Cygni-like emission profiles consistent with our HST observations. Even though \emph{l}~Car appears to have a relatively weak wind, it is adding to the picture of Cepheid mass loss in general and hints at the important connection between pulsation, convection and mass loss in these stars.

 Our measured value of the rate of period change appears to approximately match predictions from stellar evolution models. But, this is arguably an exception instead of the norm.  \cite{Neilson2012a} found that the observed rate of period change for Polaris is completely inconsistent with standard stellar evolution models and similar results were found for the population of Galactic Cepheids \citep{Neilson2012b}.  Similar issues have been found for modeling the period change in hotter $\beta$~Cephei variables, where standard stellar evolution failed to agree with measurements of period change for five stars, but agreed for three \citep{Neilson2015}.  Likewise, the distribution of rates of period change in RR~Lyrae stars are inconsistent with expectations from theoretical models \citep{LeBorgne2007}.  It is becoming clear that stellar evolution models are not yet consistent with period change observations for most cases.

One such challenge for resolving this issue is the role of metallicity.  While it is known that for a given mass, stellar winds and convective core overshooting can significantly change predicted rates of period change for a given crossing of the instability.  However, it is not clear how rates of period change vary as a function of metallicity for a given Cepheid. One would expect metallicity to matter because the width and shape of Cepheid blue loops change with metallicity. For a population of Cepheids, we cannot say how period change varies as a function of metallicity.  For instance, if metallicity affects period change in a manner similar to convective core overshooting then for a population of Cepheids varying metallicity will have a negligible impact.  We will explore this question in more detail in future work.

It is not clear if convection and convective cell eddies play a role in the rate of period change for \emph{l}~Car. \cite{Neilson2014} hypothesized that the period jitter observed in a few short-period~Cepheids \citep{Derekas2012, Evans2015} is due to changes in flux caused by large convection cells.  However, this type of period jitter has not been detected in long-period~Cepheids, although \cite{Percy2014} presented measurements of amplitude variations that might be consistent with large-scale convection cells.  Understanding the role of surface convection in long-period~Cepheids and Cepheids near the red edge of the instability strip is still an outstanding problem in stellar physics \citep{Gastine2011, Mundprecht2013, Mundprecht2015}.  Measurements of period change can provide an additional constraint of the coupling between convection and pulsation in classical Cepheids

Thanks to decades of  pulsation period measurements, we are beginning to observe stellar evolution directly in classical Cepheids. By comparing these rates of period change with stellar models, we are constraining the stellar properties and physics with great precision that will help us calibrate the Cepheid Leavitt law and understand how these Cepheids will eventually evolve into red giant and supergiant stars and beyond to white dwarf stars or explode as supernovae. For the specific case of \emph{l}~Carinae, we are observing a massive star evolving through its final crossing as a Cepheid that will cease pulsating in the next 31,000 years.

\acknowledgements
We would like to thank the anonymous referee for his/her helpful comments that have improved this work.
HN acknowledges research funding from the University of Toronto.
The authors wish to gratefully acknowledge the support from
NASA grants HST-GO11726, HST-GO12302, HST-GO13019 and NASA XMM-Newton
grant NNX14AF12G. We also thank Kyle Conroy (Vanderbilt University) for his help
with the use and improvements of the KEPHEM  program used to impersonally 
determine the times of maximum light from the photometry.
Support for HST programs 11726, 12302 and 13019 was provided by NASA through
grants from the Space Telescope Science Institute, which is operated by the Association of
Universities for Research in Astronomy, Inc., under NASA contract NAS 5-26555.

Facilities: XMM-Newton, Chandra, HST (COS) , IUE.

\bibliographystyle{apj}
\bibliography{lcar}

\begin{thebibliography}{58}
\expandafter\ifx\csname natexlab\endcsname\relax\def\natexlab#1{#1}\fi

\bibitem[{{Anderson} {et~al.}(2014){Anderson}, {Ekstr{\"o}m}, {Georgy},
  {Meynet}, {Mowlavi}, \& {Eyer}}]{Anderson2014}
{Anderson}, R.~I., {Ekstr{\"o}m}, S., {Georgy}, C., {Meynet}, G., {Mowlavi},
  N., \& {Eyer}, L. 2014, \aap, 564, A100

\bibitem[{{Anderson} {et~al.}(2015){Anderson}, {M{\'e}rand}, {Kervella},
  {Breitfelder}, {LeBouquin}, {Eyer}, {Gallenne}, {Palaversa}, {Semaan},
  {Saesen}, \& {Mowlavi}}]{Anderson2015}
{Anderson}, R.~I., {M{\'e}rand}, A., {Kervella}, P., {Breitfelder}, J.,
  {LeBouquin}, J.-B., {Eyer}, L., {Gallenne}, A., {Palaversa}, L., {Semaan},
  T., {Saesen}, S., \& {Mowlavi}, N. 2015, ArXiv e-prints

\bibitem[{{Benedict} {et~al.}(2007){Benedict}, {McArthur}, {Feast}, {Barnes},
  {Harrison}, {Patterson}, {Menzies}, {Bean}, \& {Freedman}}]{Benedict2007}
{Benedict}, G.~F., {McArthur}, B.~E., {Feast}, M.~W., {Barnes}, T.~G.,
  {Harrison}, T.~E., {Patterson}, R.~J., {Menzies}, J.~W., {Bean}, J.~L., \&
  {Freedman}, W.~L. 2007, \aj, 133, 1810

\bibitem[{{Berdnikov} {et~al.}(2003){Berdnikov}, {Mattei}, \&
  {Beck}}]{Berdnikov2003}
{Berdnikov}, L., {Mattei}, J.~A., \& {Beck}, S.~J. 2003, Journal of the
  American Association of Variable Star Observers (JAAVSO), 31, 146

\bibitem[{{Bohm-Vitense} \& {Love}(1994)}]{Vitense1994}
{Bohm-Vitense}, E. \& {Love}, S.~G. 1994, \apj, 420, 401

\bibitem[{{Bono} {et~al.}(2000){Bono}, {Castellani}, \& {Marconi}}]{Bono2000}
{Bono}, G., {Castellani}, V., \& {Marconi}, M. 2000, \apj, 529, 293

\bibitem[{{Breitfelder} {et~al.}(2016){Breitfelder}, {M{\'e}rand}, {Kervella},
  {Gallenne}, {Szabados}, {Anderson}, \& {Le Bouquin}}]{Breitfelder2016}
{Breitfelder}, J., {M{\'e}rand}, A., {Kervella}, P., {Gallenne}, A.,
  {Szabados}, L., {Anderson}, R.~I., \& {Le Bouquin}, J.-B. 2016, ArXiv
  e-prints

\bibitem[{{Caputo} {et~al.}(2005){Caputo}, {Bono}, {Fiorentino}, {Marconi}, \&
  {Musella}}]{Caputo2005}
{Caputo}, F., {Bono}, G., {Fiorentino}, G., {Marconi}, M., \& {Musella}, I.
  2005, \apj, 629, 1021

\bibitem[{{Cassisi} \& {Salaris}(2011)}]{Cassisi2011}
{Cassisi}, S. \& {Salaris}, M. 2011, \apjl, 728, L43

\bibitem[{{Cogan} {et~al.}(1980){Cogan}, {Faulkner}, \& {Butler}}]{Cogan1980}
{Cogan}, B.~C., {Faulkner}, D.~G., \& {Butler}, S.~J. 1980, \aap, 86, 283

\bibitem[{{Cousins}(1924)}]{Cousins1924}
{Cousins}, A.~W.~J. 1924, \mnras, 84, 620

\bibitem[{{Derekas} {et~al.}(2012){Derekas}, {Szab{\'o}}, {Berdnikov},
  {Szab{\'o}}, {Smolec}, {Kiss}, {Szabados}, {Chadid}, {Evans}, {Kinemuchi},
  {Nemec}, {Seader}, {Smith}, \& {Tenenbaum}}]{Derekas2012}
{Derekas}, A., {Szab{\'o}}, G.~M., {Berdnikov}, L., {Szab{\'o}}, R., {Smolec},
  R., {Kiss}, L.~L., {Szabados}, L., {Chadid}, M., {Evans}, N.~R., {Kinemuchi},
  K., {Nemec}, J.~M., {Seader}, S.~E., {Smith}, J.~C., \& {Tenenbaum}, P. 2012,
  \mnras, 425, 1312

\bibitem[{{Eddington}(1919)}]{Eddington1918}
{Eddington}, A.~S. 1919, The Observatory, 42, 338

\bibitem[{{Eldridge} \& {Tout}(2004)}]{Eldridge2004}
{Eldridge}, J.~J. \& {Tout}, C.~A. 2004, \memsai, 75, 694

\bibitem[{{Engle}(2015)}]{Engle2015}
{Engle}, S.~G. 2015, PhD thesis, James Cook University

\bibitem[{{Engle} \& {Guinan}(2012)}]{Engle2012}
{Engle}, S.~G. \& {Guinan}, E.~F. 2012, Journal of Astronomy and Space
  Sciences, 29, 181

\bibitem[{{Engle} {et~al.}(2014){Engle}, {Guinan}, {Harper}, {Neilson}, \&
  {Remage Evans}}]{Engle2014}
{Engle}, S.~G., {Guinan}, E.~F., {Harper}, G.~M., {Neilson}, H.~R., \& {Remage
  Evans}, N. 2014, \apj, 794, 80

\bibitem[{{Evans} {et~al.}(2016){Evans}, {Pillitteri}, {Wolk}, {Karovska},
  {Tingle}, {Guinan}, {Engle}, {Bond}, {Schaefer}, \& {Mason}}]{Evans2016}
{Evans}, N.~R., {Pillitteri}, I., {Wolk}, S., {Karovska}, M., {Tingle}, E.,
  {Guinan}, E., {Engle}, S., {Bond}, H.~E., {Schaefer}, G.~H., \& {Mason},
  B.~D. 2016, ArXiv e-prints

\bibitem[{{Evans} {et~al.}(2015){Evans}, {Szab{\'o}}, {Derekas}, {Szabados},
  {Cameron}, {Matthews}, {Sasselov}, {Kuschnig}, {Rowe}, {Guenther}, {Moffat},
  {Rucinski}, \& {Weiss}}]{Evans2015}
{Evans}, N.~R., {Szab{\'o}}, R., {Derekas}, A., {Szabados}, L., {Cameron}, C.,
  {Matthews}, J.~M., {Sasselov}, D., {Kuschnig}, R., {Rowe}, J.~F., {Guenther},
  D.~B., {Moffat}, A.~F.~J., {Rucinski}, S.~M., \& {Weiss}, W.~W. 2015, \mnras,
  446, 4008

\bibitem[{{Fadeyev}(2015)}]{Yel2015}
{Fadeyev}, Y.~A. 2015, \mnras, 449, 1011

\bibitem[{{Feinstein} \& {Muzzio}(1969)}]{Feinstein1969}
{Feinstein}, A. \& {Muzzio}, J.~C. 1969, \aap, 3, 388

\bibitem[{{Freedman} {et~al.}(2012){Freedman}, {Madore}, {Scowcroft}, {Burns},
  {Monson}, {Persson}, {Seibert}, \& {Rigby}}]{Freedman2012}
{Freedman}, W.~L., {Madore}, B.~F., {Scowcroft}, V., {Burns}, C., {Monson}, A.,
  {Persson}, S.~E., {Seibert}, M., \& {Rigby}, J. 2012, \apj, 758, 24

\bibitem[{{Fricke} {et~al.}(1972){Fricke}, {Stobie}, \&
  {Strittmatter}}]{Fricke1972}
{Fricke}, K., {Stobie}, R.~S., \& {Strittmatter}, P.~A. 1972, \apj, 171, 593

\bibitem[{{Gastine} \& {Dintrans}(2011)}]{Gastine2011}
{Gastine}, T. \& {Dintrans}, B. 2011, in SF2A-2011: Proceedings of the Annual
  meeting of the French Society of Astronomy and Astrophysics, ed.
  G.~{Alecian}, K.~{Belkacem}, R.~{Samadi}, \& D.~{Valls-Gabaud}, 215--219

\bibitem[{{Georgy}(2012)}]{Georgy2012}
{Georgy}, C. 2012, \aap, 538, L8

\bibitem[{{Gieren}(1989)}]{Gieren1989}
{Gieren}, W.~P. 1989, \aap, 225, 381

\bibitem[{{Innes}(1897)}]{Innes1897}
{Innes}, R.~T.~A. 1897, \aj, 17, 95

\bibitem[{{Kervella} {et~al.}(2004){Kervella}, {Fouqu{\'e}}, {Storm}, {Gieren},
  {Bersier}, {Mourard}, {Nardetto}, \& {du Coud{\'e} Foresto}}]{Kervella2004a}
{Kervella}, P., {Fouqu{\'e}}, P., {Storm}, J., {Gieren}, W.~P., {Bersier}, D.,
  {Mourard}, D., {Nardetto}, N., \& {du Coud{\'e} Foresto}, V. 2004, \apjl,
  604, L113

\bibitem[{{Kervella} {et~al.}(2009){Kervella}, {M{\'e}rand}, \&
  {Gallenne}}]{Kervella2009}
{Kervella}, P., {M{\'e}rand}, A., \& {Gallenne}, A. 2009, \aap, 498, 425

\bibitem[{{Koncewicz} \& {Jordan}(2007)}]{Koncewicz2007}
{Koncewicz}, R. \& {Jordan}, C. 2007, \mnras, 374, 220

\bibitem[{{Langer}(2012)}]{Langer2012}
{Langer}, N. 2012, \araa, 50, 107

\bibitem[{{Le Borgne} {et~al.}(2007){Le Borgne}, {Paschke}, {Vandenbroere},
  {Poretti}, {Klotz}, {Bo{\"e}r}, {Damerdji}, {Martignoni}, \&
  {Acerbi}}]{LeBorgne2007}
{Le Borgne}, J.~F., {Paschke}, A., {Vandenbroere}, J., {Poretti}, E., {Klotz},
  A., {Bo{\"e}r}, M., {Damerdji}, Y., {Martignoni}, M., \& {Acerbi}, F. 2007,
  \aap, 476, 307

\bibitem[{{Leavitt}(1908)}]{Leavitt1908}
{Leavitt}, H.~S. 1908, Annals of Harvard College Observatory, 60, 87

\bibitem[{{Luck} {et~al.}(2008){Luck}, {Andrievsky}, {Fokin}, \&
  {Kovtyukh}}]{Luck2008}
{Luck}, R.~E., {Andrievsky}, S.~M., {Fokin}, A., \& {Kovtyukh}, V.~V. 2008,
  \aj, 136, 98

\bibitem[{{Mackey} {et~al.}(2012){Mackey}, {Mohamed}, {Neilson}, {Langer}, \&
  {Meyer}}]{Mackey2012}
{Mackey}, J., {Mohamed}, S., {Neilson}, H.~R., {Langer}, N., \& {Meyer},
  D.~M.-A. 2012, \apjl, 751, L10

\bibitem[{{Maeder} \& {Meynet}(2000)}]{Maeder2000}
{Maeder}, A. \& {Meynet}, G. 2000, \araa, 38, 143

\bibitem[{{Marengo} {et~al.}(2010){Marengo}, {Evans}, {Barmby}, {Matthews},
  {Bono}, {Welch}, {Romaniello}, {Huelsman}, {Su}, \& {Fazio}}]{Marengo2010}
{Marengo}, M., {Evans}, N.~R., {Barmby}, P., {Matthews}, L.~D., {Bono}, G.,
  {Welch}, D.~L., {Romaniello}, M., {Huelsman}, D., {Su}, K.~Y.~L., \& {Fazio},
  G.~G. 2010, \apj, 725, 2392

\bibitem[{{Matthews} {et~al.}(2012){Matthews}, {Marengo}, {Evans}, \&
  {Bono}}]{Matthews2012}
{Matthews}, L.~D., {Marengo}, M., {Evans}, N.~R., \& {Bono}, G. 2012, \apj,
  744, 53

\bibitem[{{M{\'e}rand} {et~al.}(2015){M{\'e}rand}, {Kervella}, {Breitfelder},
  {Gallenne}, {Coud{\'e} du Foresto}, {ten Brummelaar}, {McAlister}, {Ridgway},
  {Sturmann}, {Sturmann}, \& {Turner}}]{Merand2015}
{M{\'e}rand}, A., {Kervella}, P., {Breitfelder}, J., {Gallenne}, A., {Coud{\'e}
  du Foresto}, V., {ten Brummelaar}, T.~A., {McAlister}, H.~A., {Ridgway}, S.,
  {Sturmann}, L., {Sturmann}, J., \& {Turner}, N.~H. 2015, \aap, 584, A80

\bibitem[{{Meyer} {et~al.}(2014){Meyer}, {Gvaramadze}, {Langer}, {Mackey},
  {Boumis}, \& {Mohamed}}]{Meyer2014}
{Meyer}, D.~M.-A., {Gvaramadze}, V.~V., {Langer}, N., {Mackey}, J., {Boumis},
  P., \& {Mohamed}, S. 2014, \mnras

\bibitem[{{Mundprecht} {et~al.}(2013){Mundprecht}, {Muthsam}, \&
  {Kupka}}]{Mundprecht2013}
{Mundprecht}, E., {Muthsam}, H.~J., \& {Kupka}, F. 2013, \mnras, 435, 3191

\bibitem[{{Mundprecht} {et~al.}(2015){Mundprecht}, {Muthsam}, \&
  {Kupka}}]{Mundprecht2015}
---. 2015, \mnras, 449, 2539

\bibitem[{{Neilson}(2014)}]{Neilson2014b}
{Neilson}, H.~R. 2014, \aap, 563, A48

\bibitem[{{Neilson} {et~al.}(2011){Neilson}, {Cantiello}, \&
  {Langer}}]{Neilson2011}
{Neilson}, H.~R., {Cantiello}, M., \& {Langer}, N. 2011, \aap, 529, L9

\bibitem[{{Neilson} {et~al.}(2012{\natexlab{a}}){Neilson}, {Engle}, {Guinan},
  {Langer}, {Wasatonic}, \& {Williams}}]{Neilson2012a}
{Neilson}, H.~R., {Engle}, S.~G., {Guinan}, E., {Langer}, N., {Wasatonic},
  R.~P., \& {Williams}, D.~B. 2012{\natexlab{a}}, \apjl, 745, L32

\bibitem[{{Neilson} \& {Ignace}(2014)}]{Neilson2014}
{Neilson}, H.~R. \& {Ignace}, R. 2014, \aap, 563, L4

\bibitem[{{Neilson} \& {Ignace}(2015)}]{Neilson2015}
---. 2015, \aap, 584, A58

\bibitem[{{Neilson} {et~al.}(2012{\natexlab{b}}){Neilson}, {Langer}, {Engle},
  {Guinan}, \& {Izzard}}]{Neilson2012b}
{Neilson}, H.~R., {Langer}, N., {Engle}, S.~G., {Guinan}, E., \& {Izzard}, R.
  2012{\natexlab{b}}, \apjl, 760, L18

\bibitem[{{Neilson} \& {Lester}(2008)}]{Neilson2008}
{Neilson}, H.~R. \& {Lester}, J.~B. 2008, \apj, 684, 569

\bibitem[{{Neilson} \& {Lester}(2009)}]{Neilson2009}
---. 2009, \apj, 690, 1829

\bibitem[{{Percy}(2007)}]{Percy2008}
{Percy}, J.~R. 2007, {Understanding Variable Stars} (Cambridge, U.K.: Cambridge
  University Press)

\bibitem[{{Percy} \& {Kim}(2014)}]{Percy2014}
{Percy}, J.~R. \& {Kim}, R.~Y.~H. 2014, Journal of the American Association of
  Variable Star Observers (JAAVSO), 42, 267

\bibitem[{{Poleski}(2008)}]{Poleski2008}
{Poleski}, R. 2008, AcA, 58, 313

\bibitem[{{Roberts}(1901)}]{Roberts1901}
{Roberts}, A.~W. 1901, \aj, 21, 81

\bibitem[{{Turner}(2010)}]{Turner2010}
{Turner}, D.~G. 2010, \apss, 326, 219

\bibitem[{{Turner} {et~al.}(2006){Turner}, {Abdel-Sabour Abdel-Latif}, \&
  {Berdnikov}}]{Turner2006}
{Turner}, D.~G., {Abdel-Sabour Abdel-Latif}, M., \& {Berdnikov}, L.~N. 2006,
  \pasp, 118, 410

\bibitem[{{Walmswell} {et~al.}(2015){Walmswell}, {Tout}, \&
  {Eldridge}}]{Eldridge2015}
{Walmswell}, J.~J., {Tout}, C.~A., \& {Eldridge}, J.~J. 2015, \mnras, 447, 2951

\bibitem[{{Yoon} \& {Langer}(2005)}]{Yoon2005}
{Yoon}, S.-C. \& {Langer}, N. 2005, \aap, 435, 967

\end{thebibliography}
\end{document}